\documentclass[a4paper,cite]{article}
\usepackage{latexsym}
\usepackage{graphicx}
\newcommand{\lbl}[1]{\label{eq:#1}}
\newcommand{ \rf}[1]{(\ref{eq:#1})}

\newcommand{\be}{\begin{equation}}
\newcommand{\ee}{\end{equation}}
\newcommand{\bea}{\begin{eqnarray}}
\newcommand{\eea}{\end{eqnarray}}
\newcommand{\setl}{\setlength\arraycolsep{2pt}}

\newcommand{\noi}{\noindent}
\newcommand{\nn}{\nonumber}
\newcommand{\ra}{\rightarrow}
\newcommand{\Ra}{\Rightarrow}

\newcommand{\lesssim}{ {\
\lower-1.2pt\vbox{\hbox{\rlap{$<$}\lower5pt\vbox{\hbox{$\sim$}}}}\ } 
}
\newcommand{\gtrsim}{ {\
\lower-1.2pt\vbox{\hbox{\rlap{$>$}\lower5pt\vbox{\hbox{$\sim$}}}}\ } 
}

\newcommand{\cD}{{\cal D}}

\newcommand{\cH}{{\cal H}}

\newcommand{\cL}{{\cal L}}

\newcommand{\cO}{{\cal O}}

\newcommand{\cR}{{\cal R}}
\newcommand{\cS}{{\cal S}}

\newcommand{\cW}{{\cal W}}

\newcommand{\Tr}{\mbox{\rm Tr}}
\newcommand{\tr}{\mbox{\rm tr}}

\newcommand{\MeV}{\mbox{\rm MeV}}
\newcommand{\GeV}{\mbox{\rm GeV}}

\newcommand{\with}{\mbox{\rm with}}

\newcommand{\annd}{\mbox{\rm and}}
\newcommand{\foor}{\mbox{\rm for}}

\newcommand{\als}{\alpha_{\mbox{\rm {\scriptsize s}}}}

\newcommand{\muhad}{\mu_{\mbox{\rm {\scriptsize had.}}}}

\newcommand{\gL}{\frac{1-\gamma_{5}}{2}}
\newcommand{\gR}{\frac{1+\gamma_{5}}{2}}

\newcommand{\QCD}{\mbox{\rm {\footnotesize QCD}}}

\newcommand{\bP}{{\bf P}}

\newcommand{\Psls}{\not \! \bP}

\input epsf

\begin{document}

\begin{titlepage}

\begin{flushright} CPT-2000/P.4015\\ UAB-FT-487\\ 
%\today
\end{flushright}
\vspace*{1.5cm}
\begin{center} 
{\Large \bf $K^{0}$--$\bar{K^{0}}$ Mixing in the $1/N_c$
Expansion\footnote{This is a revised version of our article in Phys. Lett.
{\bf B490}
  (2000) 213.}}\\[3.0cm]

{\bf Santiago Peris}$^a$ and {\bf Eduardo de Rafael}$^b$\\[1cm]

$^a$ Grup de F{\'\i}sica Te{\`o}rica and IFAE\\ Universitat Aut{\`o}noma
de Barcelona, 08193 Barcelona, Spain.\\[0.5cm]
$^b$  Centre  de Physique Th{\'e}orique\\
       CNRS-Luminy, Case 907\\
    F-13288 Marseille Cedex 9, France\\

\end{center}

\vspace*{1.0cm}

\begin{abstract}

We present the result for the invariant $\hat{B}_{K}$ factor of
$K^{0}$--$\bar{K^{0}}$ mixing in the chiral limit and to next--to--
leading order in the
$1/N_c$ expansion. We explicitly demonstrate the cancellation of the
renormalization scale and scheme dependences between short-- and
long--distance contributions in the final expression. Numerical estimates 
are then given, by taking into account increasingly refined short-- 
and long--distance constraints of the underlying QCD Green's function which
governs  the $\hat{B}_{K}$ factor.      
\end{abstract}

\end{titlepage}

\section{Introduction}
\lbl{sec:int}

\noi
The Standard Model predicts strangeness changing transitions with
$\Delta S=2$ via two virtual $W$--exchanges between quark lines, 
the so--called box diagrams. The low--energy physics of
these transitions is governed by an effective Hamiltonian which is 
proportional to the local four--quark operator, (summation over colour indices
within brackets is understood and $q_{L}\equiv\gL q$,) 
\be
\label{eq:deltas2} Q_{\Delta S=2}(x)\equiv
(\bar{s}_{L}(x)\gamma^{\mu}d_{L}(x))(\bar{s}_{L}(x)\gamma_{\mu}d_{L}(x)),
\ee
modulated by a quadratic form of the flavour mixing matrix elements
$\lambda_{q}=V_{\rm qd}^{\ast}V_{\rm qs}\,,\,{\rm q}={\rm u,c,t}\,,$
with coefficient functions $F_{1,2,3}$ of the heavy masses  of the 
fields $t$, $Z^0$, $W^{\pm}$, $b$, and $c$ which have been integrated
out~\footnote{For a detailed discussion, see e.g. refs.~\cite{BBL96,Bu99}
and references therein.}:
\be\lbl{effhal} 
{\cH}_{\rm eff}^{\Delta S=2}=\frac{G_{F}^{2}M_{W}^2}{4\pi^2}
\left[\lambda_{c}^2F_1+\lambda_{t}^2F_2+2\lambda_{c}\lambda_{t}F_3\right]
C_{\Delta S=2}(\mu)Q_{\Delta S=2}(x)\, .
\ee
The operator $Q_{\Delta S=2}$ is multiplicatively renormalizable 
and has
an anomalous dimension $\gamma(\alpha_{s})$ which in perturbative QCD (pQCD)
is  defined by the equation
\be
\mu^2\frac{d}{d\mu^2}<Q_{\Delta S=2}>=-\frac{1}{2}\gamma(\alpha_{s}) 
<Q_{\Delta
S=2}>\,,\quad \gamma(\alpha_s)=\frac{\als}{\pi}\gamma_1 +
\left(\frac{\als}{\pi}\right)^2\gamma_2+\cdots\,.
\ee
with $\left(\beta_{1}=\frac{1}{6}(-11N_c +2n_f)\right)$ 
\be
\gamma_1= \frac{3}{2}\left(1-\frac{1}{N_c}\right)\quad\annd\quad
\gamma_2=\frac{1}{32}\left(1-\frac{1}{N_c}\right)\left[-\frac{19}{3}N_c+
\frac{4}{3}n_f -21 +57\frac{1}{N_c}+4\beta_{1}\kappa\right]\,,
\ee
where $\kappa\!=\!0$ in the {\it na\"{\i}ve dimensional renormalization 
scheme}  (NDR) and
$\kappa\!=\!-4$ in the {\it 't Hooft--Veltman renormalization scheme} (HV). The
renormalization $\mu$--scale dependence of the Wilson coefficient
$C_{\Delta S=2}(\mu)$ in Eq.~\rf{effhal} is
then, 
$\left(\beta_{2}=-\frac{17}{12}N_c^2+\frac{1}{4}\frac{N_c^2-1}{2N_C}n_f+
\frac{5}{12}N_cn_f\right)\,,$
{\setl
\bea
C_{\Delta
S=2}(\mu)
& = &
\left[1+\frac{\als(\mu)}{\pi}\frac{1}{\beta_{1}}\left(\gamma_2+
\frac{\gamma_1}{-\beta_1}\beta_{2}
\right)\right]
\left(\frac{1}{\als(\mu)}\right)^{\frac{\gamma_1}{-\beta_1}}
 \\ \lbl{wilsonN}
 & \ra & 
\left[1+\frac{\als(\mu)}{\pi}\left(\frac{1433}{1936}+\frac{1}{8}\kappa
\right)\right]
\left(\frac{1}{\als(\mu)}\right)^{\frac{9}{11}\frac{1}N_c}\,,
\eea}

\noi
where the second line gives the result to next--to--leading order in the
$1/N_c$ expansion, which is the approximation at which we shall be 
working here.

The matrix element
\be
\label{eq:Bpar} <\bar{K}^0|Q_{\Delta S=2}(0)|K^0>\equiv
\frac{4}{3}f_{K}^2M_{K}^2B_{K}(\mu),
\ee
defines the so--called $B_{K}$--parameter of 
$K^0-\bar{K}^0$
mixing at short--distances, which is one of the crucial parameters in the
phenomenological studies of CP--violation in the Standard Model. 
In the large--$N_c$ limit of QCD, the  four quark operator $Q_{\Delta S=2}$ 
factorizes
into a product of two current operators. 
Each of these
currents, to lowest order in chiral perturbation theory, 
has a simple bosonic realization:
\be 
(\bar{s}_L\gamma^{\mu}d_L)=\frac{\delta\cL_{\QCD}(x)}{\delta
l_{\mu}(x)_{32}}\Ra
\frac{i}{2}F_{0}^2\tr\left[\lambda_{32}(D^{\mu}U^{\dagger})U\right]=
\frac{1}{\sqrt{2}}F_{0}
\partial^{\mu}K^{0}
+\cdots\,,
\ee
where $F_{0}$ denotes the coupling constant of the pion in the
chiral limit and 
$U$ is the $3\times 3$ unitary matrix in flavour space, $U=1+\cdots$, which
collects the Goldstone fields and which under chiral rotations 
transforms like $U\ra V_{R}UV_{L}^{\dagger}$. The large--$N_c$ 
approximation, with inclusion of the chiral corrections in the
factorized contribution,  leads to the result:
\be\lbl{BKlargeN} 
B_{K}|_{N_c\ra \infty}=\frac{3}{4}\quad\annd\quad C_{\Delta S=2}
(\mu)|_{N_c\ra\infty}= 1\,.
\ee
In full generality, the bosonization of the four--quark operator $Q_{\Delta
S=2}(x)$ to lowest order in the chiral expansion is described by an
effective operator which is of
$\cO(p^2)$~\footnote{See e.g. refs.~\cite{PdeR91,deR95}}:
\be\lbl{CHI2}
Q_{\Delta S=2}(x)\Ra -\frac{F_{0}^4}{4}\ g_{\Delta
S=2}(\mu)\ 
\tr\left[
\lambda_{32}(D^{\mu}U^{\dagger})U\lambda_{32}(D_{\mu}U^{\dagger})U\right]
\,,
\ee
where $\lambda_{32}$ denotes the matrix $\lambda_{32}=\delta_{i3}
\delta_{2j}$ in flavour space and $g_{\Delta
S=2}(\mu)$ is a dimensionless coupling constant which depends on the 
underlying dynamics of spontaneous chiral symmetry breaking (S$\chi$SB) in
QCD.  The relation between the coupling $g_{\Delta S=2}(\mu)$ and the
phenomenological
$B_{K}$ factor defined in Eq.~\rf{Bpar} is
simply
\be
B_{K}(\mu,m_{u,d,s}\ra 0)=\frac{3}{4}g_{\Delta S=2}(\mu)\,.
\ee
Notice that the coupling
$g_{\Delta S=2}(\mu)$, and hence $B_{K}$, is perfectly well defined in 
the chiral limit, while the matrix element in Eq.~\rf{Bpar} vanishes in the
chiral limit as a chiral power. To lowest order in the chiral
expansion, the relation  to the so called {\it invariant}
$\hat{B}_{K}$--factor is as follows:
\be
\hat{B}_{K}=\frac{3}{4}C_{\Delta S=2}(\mu)\times g_{\Delta S=2}(\mu)\,,
\ee
which means that the coupling  $g_{\Delta S=2}(\mu)$ must have a
$\mu$ dependence (and a scheme dependence) which must cancel
with the
$\mu$ dependence  (and scheme dependence) of the Wilson coefficient
$C_{\Delta S=2}(\mu)$. The purpose of this note is to show how the 
mechanism of cancellations works in practice within the framework of the
$1/N_c$ expansion. This will allow us to give a numerical result for
$\hat{B}_{K}$ which is valid to lowest order in the chiral expansion and to
next--to--leading order in the
$1/N_c$ expansion. We wish to emphasize that the novel feature of this work is
that, to our knowledge, it is the first calculation  of $\hat{B}_K$ which
explicitly shows the cancellations of scale  and scheme dependences.
%%%%%%%%%%%%%%%%%%%%%%%%%%%%%%%%%%%%%%%%%%%

\section{Bosonization of Four--Quark Operators}
\setcounter{equation}{0}
\lbl{sec:bosfqo} 

\noi
The QCD Lagrangian in the presence of external chiral sources $l_{\mu}$,
$r_{\mu}$ of left-- and right-- currents, but with neglect of scalar and
pseudoscalar sources which is justified in the chiral limit approximation
we shall be working here, has the form
\be
\cL_{\QCD}=-\frac{1}{4}G_{\mu\nu}^{(a)}G^{(a)\mu\nu}+i\bar{q}\cD_{\chi}q\,,
\ee
with $\cD_{\chi}$ the Dirac operator
\be
\cD_{\chi} 
  =  \gamma^{\mu}(\partial_{\mu}+iG_{\mu})-
i\gamma^{\mu}\left[l_{\mu}\gL+r_{\mu}\gR\right]
\,. 
\ee
The bosonization of the four--quark operator $Q_{\Delta S=2}(x)$ is
formally defined by the functional integral~\cite{PdeR91}
\bea\lbl{S=2}
\lefteqn{\langle Q_{\Delta S=2}(x)\rangle=\Tr\ \cD_{\chi}^{-1}
i\frac{\delta\cD_{\chi}}{\delta l_{\mu}(x)_{32}}\Tr\ \cD_{\chi}^{-1}
i\frac{\delta\cD_{\chi}}{\delta l^{\mu}(x)_{32}}} \nn \\ & & -
\int d^4 y\int\frac{d^4 q}{(2\pi)^4}e^{-iq\cdot (x-y)}\ \Tr
\left(\cD_{\chi}^{-1}i\frac{\delta\cD_{\chi}}{\delta
l_{\mu}(x)_{32}}\cD_{\chi}^{-1} i\frac{\delta\cD_{\chi}}{\delta
l^{\mu}(y)_{32}} \right)\,,
\eea
where the trace $\Tr$ here also includes the
functional integration over gluons in large $N_c$.
The first term corresponds to the factorized pattern and gives the
contributions of $\cO(N_c^2)$; the second term corresponds to the
unfactorized pattern and it involves an integral, (which is 
{\it regularization} dependent,) over all virtual momenta $q$. This is the term
which gives  the next--to--leading
$\cO(N_c)$ contribution we are interested in.

To proceed further, it is convenient to use the Schwinger's operator
formalism. With 
$\Psls$ the conjugate momentum operator, in the absence of external chiral
fields, the full quark propagator is 
\be
(x\vert \frac{1}{\cD_{\chi}}\vert y)=
(x\vert
\frac{i}{\Psls+\gamma^{\alpha}\left[l_{\alpha}\gL+r_{\alpha}\gR\right]}
\vert
y)\quad\with\quad (x\vert \Psls\vert
y)=\gamma^{\mu}\left[i\frac{\partial}{\partial_{\mu}}
-G_{\mu}\right]\delta(x-y)\,.
\ee
The chiral expansion is then defined as an expansion in the $l_{\alpha}$ 
and $r_{\alpha}$ external sources.
The precise relation between the formal bosonization in
Eq.~\rf{S=2} and the explicit chiral realization in Eq.~\rf{CHI2}
can best be seen from the fact that:
$D^{\mu}U^{\dagger}U=\partial^{\mu}U^{\dagger}U+iU^{\dagger}
r^{\mu}U -il^{\mu}$. This shows that there are a priori
{\it six} different ways to compute the constant $g_{\Delta S=2}(\mu)$,
although they are all related by chiral gauge invariance. One possible
choice is the term~
\footnote{Other
choices are indeed possible but, in general, the underlying QCD
Green's functions have pieces which are not order parameters and spurious
contributions which depend on the regularization. We have shown 
the equivalence among the various possible choices, but we postpone the
detailed discussion which is rather technical to a longer publication.}
\be
\cL_{\underline{27}}^{\Delta S=2}(x) = -\frac{F_{0}^4}{4}\
g_{\Delta S=2}(\mu)\left\{\cdots 
-\tr[\lambda_{32}U^{\dagger}r^{\mu}U
\lambda_{32} U^{\dagger}r_{\mu}U]+  \cdots\right\}
\,.
\ee 
This is a convenient choice because the underlying QCD Green's function is
the four--point function, [
$L_{\bar{s}d}^{\mu}(x)\equiv\sum_{a}\bar{s}^{a}(x)\gamma^{\mu}\gL d^{a}(x)$ and
$R_{\bar{d}s}^{\mu}(x)\equiv\sum_{a}\bar{d}^{a}(x)\gamma^{\mu}\gR s^{a}(x)$,]
\be\lbl{LRLR}
\cW_{LRLR}^{\mu\alpha\nu\beta}(q,l)=\lim_{l\ra 0}\ i^3 \int d^4x\ d^4y\
d^4z\ e^{iq\cdot x}\ e^{il\cdot y}\ e^{-il\cdot z}\langle 0\vert T\{
L_{\bar{s}d}^{\mu}(x)\ R_{\bar{d}s}^{\alpha}(y)
\ L_{\bar{s}d}^{\nu}(0)\ R_{\bar{d}s}^{\beta}(z)\}\vert 0\rangle\,,
\ee 
In fact, what we need, as seen in Eq.~\rf{S=2}, is the integral of the
unfactorized four--point function $\cW_{LRLR}^{\mu\alpha\nu\beta}(q,l)$
over the  four--vector $q$ with the Lorentz indices of the two
left--currents contracted. This is a quantity which is a good order
parameter of S$\chi$SB. The integral over the solid angle has the form,
($Q^2\equiv -q^2$,)
\be\lbl{intLRLR}
\int
d\Omega_{q}\
g_{\mu\nu}\cW_{LRLR}^{\mu\alpha\nu\beta}(q,l)\vert_{\mbox{\rm
{\scriptsize unfactorized}}}=
\left(\frac{l^{\alpha}l^{\beta}}{ l^2}-g^{\alpha\beta}\right)\
\cW^{(1)}_{LRLR}(Q^2)\,,
\ee
where the transversality in the four--vector
$l$ follows from current algebra Ward identities. We are still left with an
integral of the invariant function $\cW^{(1)}_{LRLR}(Q^2)$  over the full
euclidean range: $0\le
Q^2\le\infty$ which has to be done in the same {\it renormalization
scheme} as the calculation of the short--distance Wilson
coefficient
$C_{\Delta S=2}(x)$ in Eq.~\rf{effhal} has been done, i.e. in the
$\overline{MS}$--{\it scheme}. The coupling constant  $g_{\Delta S=2}(\mu)$ defined
with {\it dimensional regularization} in $d=4-\epsilon$ dimensions is then
given by the following integral,
\be\lbl{GDELTAS2}
 g_{\Delta S=2}(\mu,\epsilon)=
1-\frac{\muhad^2}{32\pi^2
F_{0}^2}\
\frac{(4\pi\mu^2/\muhad^2)^{\epsilon/2}}{\Gamma(2-\epsilon/2)}
\int_{0}^{\infty}
dz
\ z^{-\epsilon/2}\left(W[z]\equiv z\frac{\muhad^2}{F_{0}^2}\cW^{(1)}_{LRLR}(z
\muhad^2)\right)\,,
\ee 
where, for convenience, we have normalized $Q^2$ to a characteristic hadronic
scale $\muhad^2$, ($z\equiv Q^2/\muhad^2$,) which in practice we shall choose
within the range:
$1\,\GeV\le\muhad\le 2\, \GeV$.
%%%%%%%%%%%%%%%%%%%%%%%%%%%%%%%%%%%%%%%%%%%
%%%%%%%%%%%%%%%%%%%%%%%%%%%%%%%%%%%%%%%%%%%

\section{$\hat{B}_{K}$ to Next--to--Leading Order in the $1/N_c$ Expansion}
\setcounter{equation}{0}
\lbl{sec:largeNeval}

\noi
In full generality, Green's functions in QCD at large $N_c$~\cite{tH74} 
are given by sums over an infinite number of hadronic poles~\cite{Wi79}. 
Since
the function $\cW^{(1)}_{LRLR}(Q^2)$ is an order parameter of spontaneous 
chiral symmetry breaking it receives no contribution from the perturbative
continuum and satisfies an unsubtracted dispersion relation. After repeated
use of partial fractions decomposition, one can see that for  
the particular case of the function
$W[z]$, [see Eqs.~\rf{GDELTAS2}, \rf{intLRLR} and \rf{LRLR}] 
the most general ansatz in large--$N_c$ QCD is an
infinite sum of poles, double poles and triple poles~\footnote{We
thank J.~Bijnens and J.~Prades for pointing out to us the existence of triple
poles, which were ignored in the previous version of this manuscript.} of the
form
\be
\lbl{largeNansatz}
W[z]=6-\sum_{i=1}^{\infty}\frac{\alpha_i}{\rho_{i}}-\sum_{i=1}^{\infty}
\frac{\beta_i}{\rho_{i}^2}-\sum_{i=1}^{\infty}\frac{\gamma_i}{\rho_{i}^3}
+\sum_{i=1}^{\infty}\left\{\frac{\alpha_i}{(z+\rho_{i})}+
\frac{\beta_i}{(z+\rho_{i})^2}+\frac{\gamma_i}{(z+\rho_{i})^3}\right\}\,,
\ee
where $\rho_{i}=M_{i}^2/\muhad^2$ and $M_{i}$ is the mass of the $i$--th
narrow hadronic state. Triple poles result from the combination of two
odd--parity couplings involving vector mesons and Goldstone particles.  The
first term on the r.h.s. is the contribution from the Goldstone poles to the
$W[z]$ function.  This constant term is known from previous calculations, (see
refs.~\cite{GF95,HKS99,BP99}) with which we agree. Constant terms,
however, do not contribute to the integral in Eq.~\rf{GDELTAS2} defined in
{\it dimensional regularization}. In the absence of a solution of large--$N_c$
QCD, the actual values of the mass ratios $\rho_{i}$, and residues
$\alpha_{i}$ , $\beta_{i}$ and $\gamma_i$ remain unknown. 
As discussed below, there are,
however, short--distance and long--distance constraints that the 
function  $W[z]$ has to obey.   With $W[z]$ given by Eq.~\rf{largeNansatz},
the integral in Eq.~\rf{GDELTAS2} can be done analytically, with the
result {\setl
\bea
g_{\Delta S=2}(\mu,\epsilon) & = & 1-\frac{\muhad^2}{32\pi^2
F_{0}^2}\left\{\left(\frac{2}{\epsilon}+ \log
4\pi-\gamma_{E}+\log\mu^2/\muhad^2+1\right)\sum_{i=1}^{\infty}\alpha_{i}
\right. 
\nn \\
 &  & \left. \ \ \ \ \ \ \ \ \ \ \ \ \ \ \ \ \ \ \
-\sum_{i=1}^{\infty}\alpha_{i}\log\rho_{i}
+\sum_{i=1}^{\infty}\frac{\beta_i}{\rho_{i}}+ 
\frac{1}{2}\sum_{i=1}^{\infty}\frac{\gamma_i}{\rho_{i}^2}\right\}\,.
\eea}

We shall next explore the short--distance properties and long--distance
properties which are pres\-ent\-ly known about the function $W[z]$, and hence
about the parameters $\alpha_{i}$, $\beta_{i}$, $\gamma_i$ and $\rho_{i}$.

\begin{itemize}

\item 
The large--$Q^2$ behaviour of the function $\cW^{(1)}_{LRLR}(Q^2)$ 
is governed by the OPE. We find the result
\be\lbl{OPE4}
\lim_{Q^2\ra\infty}\cW^{(1)}_{LRLR}(Q^2)=24\pi^2\frac{\als}{\pi}\left[1+
\frac{\epsilon}{2}\frac{1}{6}(5+\kappa)+\cO\left(\frac{\als}{\pi}\right)^2
\right]\frac{F_{0}^4}{Q^4}+ \dots \,,
\ee
with $\kappa=0$ in the NDR {\it scheme} and $\kappa=-4$ in the HV {\it
scheme}. The fact that the residue of the $\frac{1}{Q^4}$ power 
term in the OPE is known, entails the constraint
\be\lbl{opeconstraint}
\sum_{i=1}^{\infty}\alpha_{i}=\cR+\frac{\epsilon}{2}\cS\,, 
\ee
with
\be
\cR=\left[24\pi^2\frac{\als}{\pi}+\cO\left(\frac{N_c
\als^{2}}{\pi^{2}}\right)\right]
\frac{ F_{0}^2}{\muhad^2}\quad\annd\quad \cS=\left[4\pi^2
(5+\kappa)\frac{\als}
{\pi}+\cO\left(\frac{N_c\als^{2}}{\pi^{2}}\right)\right]
\frac{F_{0}^2}{\muhad^2}\,.
\ee
This allows us to define the renormalized coupling constant $g_{\Delta
S=2}^{(r)}(\mu)$ in the $\overline{MS}$--{\it scheme}:
\bea
\lefteqn{g_{\Delta
S=2}^{(r)}(\mu)  =  1-\frac{\muhad^2}{32\pi^2
F_{0}^2}\left\{\cR
\log\mu^2/\muhad^2+\cR+\cS+\sum_{i=1}^{\infty}\left[-\alpha_{i}\log\rho_{i}+
\frac{\beta_i}{\rho_{i}} + \frac{1}{2}\frac{\gamma_i}{\rho_i^2}\right] 
\right\}} \\
& \Ra & \left(\frac{\als(\mu)}{\als(\muhad)}\right)^{\frac{9}{11}\frac{1}{N_c}}
\left\{1-\frac{1}{2}\frac{\muhad^2}{16\pi^2
F_{0}^2}\left(\cR +\cS +
\sum_{i=1}^{\infty}\left[-\alpha_{i}\log\rho_{i}+
\frac{\beta_i}{\rho_{i}} +\frac{1}{2}\frac{\gamma_i}{\rho_i^2}\right]
\right) \right\} \,,
\eea
where in the second line we have written the renormalization group improved
result. We insist on the fact that this result, contrary to the various
large--$N_c$  inspired calculations which have been published so
far~\cite{BBG88,G90,PdeR91,GF95,BEFL98,HKS99,BP99} is the  
\underline {full} result  to next--to--leading order in the $1/N_c$
expansion. The renormalization $\mu$ scale dependence as well as the
scheme dependence in the factor
$\cS$ cancel exactly, at the next--to--leading log approximation, 
with the short distance $\mu$ and scheme dependences in the Wilson
coefficient
$C_{\Delta S=2}(\mu)$ in Eq.~\rf{wilsonN}. Therefore, the \underline{full}
expression of the invariant
$\hat{B}_{K}$ factor to lowest order in the chiral expansion and to
next--to--leading order in the $1/N_c$ expansion  which takes 
into account the hadronic contribution from light quarks below a mass scale
$\muhad$ is then  
{\setl\bea\lbl{finalBK}
\hat{B}_{K} & = &
\left(\frac{1}{\als(\muhad)}\right)^{\frac{3}{11}}\frac{3}{4}\left\{1  - 
\frac{\als(\muhad)}{\pi}\frac{1229}{1936}+\cO\left(\frac{N_c
\als^{2}(\muhad)}{\pi^{2}}\right)\right. \nn \\
 &  &\ \ \ \ \ \ \ \ \ \ \ \ \ \ \ \ \ \ \ \ \ \ \ \ \ \  -\left.
\frac{\muhad^2}
{32\pi^2
F_{0}^2}
\sum_{i=1}^{\infty}\left[-\alpha_{i}\log\rho_{i}+
\frac{\beta_i}{\rho_{i}}+\frac{1}{2}\frac{\gamma_i}{\rho_i^2}\right]
\right\}\,.
\eea}

\noi
The numerical choice of the hadronic scale $\muhad$ is, a priori, arbitrary.
In practice, however, $\muhad$ has to be sufficiently large so as to make
meaningful the truncated pQCD series in the first line. Our final error in the
numerical evaluation of
$\hat{B}_{K}$ will include the small effect of fixing $\muhad$ within the
range $1\, \GeV\le\muhad\le 3\, \GeV$.   

\item
The fact that there is no $1/Q^2$ term in the OPE expansion of the function
$\cW^{(1)}_{LRLR}(Q^2)$ implies the sum rule
\be\lbl{Q2}
\sum_{i=1}^{\infty}\frac{\alpha_i}{\rho_{i}}+\sum_{i=1}^{\infty}\frac{\beta_i}
{\rho_{i}^2}+ \sum_{i=1}^{\infty}\frac{\gamma_i}{\rho_{i}^3}=6\,.
\ee
This sum rule plays an equivalent r\^{o}le to the 1st Weinberg 
sum rule in the case of the
$\Pi_{LR}(Q^2)$ two--point function~\footnote{See e.g. ref.\cite{KPdeR98}}. 
Like in this case, it guarantees the cancellation of quadratic 
divergences in the integral in Eq.~\rf{GDELTAS2} and it relates the
contribution from Goldstone particles to a specific sum of resonance state
contributions. What this means in practice is that, although the Goldstone
pole does not contribute to the integral in Eq.~\rf{GDELTAS2} defined in
{\it dimensional regularization}, the {\it normalization} of the function
$W[z]$ at
$z=0$ is, precisely, fixed by the {\it residue} of the Goldstone pole
contribution; i.e., $W[0]=6$.
%
%\item
%We have found that there is no $1/Q^6$ power correction in the 
%OPE expansion of the function $\cW^{(1)}_{LRLR}(Q^2)$, which implies the
%additional short--distance sum rule
%\be\lbl{Q6}
%\sum_{i=1}^{\infty}\alpha_{i}\rho_{i}=\sum_{i=1}^{\infty}\beta_{i}\,.
%\ee

\item
The slope at the origin of the function $W[z]$ is fixed by a linear
combination of coupling constants of the $\cO(p^4)$ Gasser--Leutwyler
Lagrangian~\cite{GL85}, with the result~\footnote{The r.h.s. 
in Eq.~\rf{slope} agrees with the expression reported in
ref.~\cite{BP99}.}
\be\lbl{slope}
\sum_{i=1}^{\infty}\frac{\alpha_i}{\rho_{i}^2}+2\sum_{i=1}^{\infty}
\frac{\beta_i}{\rho_{i}^3}+ 3 
\sum_{i=1}^{\infty}\frac{\gamma_i}{\rho_{i}^4} =24\frac{\muhad^2}
{F_{0}^2}[2L_{1}+5L_{2}+L_{3}+L_{9}]\,.
\ee

\end{itemize}

The numerical evaluation of $\hat{B}_{K}$ in Eq.~\rf{finalBK} requires
the input of the mass ratios $\rho_{i}$ and couplings $\alpha_{i}$, 
$\beta_{i}$ and $\gamma_i$ 
of the narrow states which fill the details of the hadronic 
spectrum of light flavours. 
Our goal is to find the {\it minimal hadronic ansatz} of pole 
terms of the form shown in Eq.~\rf{largeNansatz} which
satisfies the short--distance and long--distance constraints discussed
above,  and use then this {\it minimal hadronic ansatz} to evaluate the
integral in Eq.~\rf{GDELTAS2}. The fact that the function
$W[z]$ is an order parameter of S$\chi$SB ensures that it has 
a smooth behaviour in the ultraviolet and it is then reasonable to
approximate it by a few states. (There is no need here for an infinite
number of narrow states to reproduce the asymptotic behaviour of
pQCD, as it would be the case with a Green's function which is
not an order parameter of S$\chi$SB.)  This procedure has been
successfully tested in other examples, like the calculation of the
electroweak
$\pi^{+}-\pi^{0}$ mass difference~\cite{KPdeR98} and the decay of 
pseudoscalars into lepton pairs~\cite{KPPdeR99}, and it has been
applied~\cite{KPdeR99} to the evaluation of matrix elements of electroweak
penguin operators as well~\footnote{See also refs.~\cite{C98,A98,LDL98}
and  refs.~\cite{DG99,N00,BP00} for other recent evaluations of electroweak
penguin operators.}. 

%%%%%%%%%%%%%%%%%%%%%%%%%%%%%%%%%%%%%%%%%%%
%%%%%%%%%%%%%%%%%%%%%%%%%%%%%%%%%%%%%%%%%%%
%%%%%%%%%%%%%%%%%%%%%%%%%%%%%%%%%%%%%%%%%%%

\section{ Numerical Evaluation and Conclusions}
\setcounter{equation}{0}
\lbl{sec:neco}

\noi
We now proceed to the numerical evaluation of the analytic
result for
$\hat{B}_{K}$ in Eq.~\rf{finalBK} by successive improvements in the 
hadronic input. 
 
The crudest approximation is the one where the only terms of the
next--to--leading order in the $1/N_c$ expansion which are retained are 
those from the first line in Eq.~\rf{finalBK}. This implicitly assumes that
all the way down to the hadronic $\muhad$ scale there is only pQCD running and
that the  hadronic contribution below the $\muhad$ scale due to light
quarks is negligible. One can see that from the integral in Eq.~\rf{GDELTAS2},
if we write it in an equivalent cut--off form:
\be
\int_{0}^{1} dz\, W[z]+\int_{1}^{\Lambda^2/\muhad^2}dz\, W_{\mbox{\rm
{\scriptsize OPE}}}[z]\,,
\ee
with $\Lambda^2=\mu^{2}\exp\left({1+\frac{\cS}{\cR}}\right)$, and where 
in the
second integral we use the asymptotic OPE expression in Eq.~\rf{OPE4}. The
approximation we are discussing is the one where the first low--energy 
integral is simply neglected.
The result of this next--to--leading log (nll) approximation is to bring up the
large--$N_c$ prediction in Eq.~\rf{BKlargeN} to a value which, including
an estimate of higher order corrections, (we use
$\Lambda_{\overline{MS}}^{(3)}=(372\pm 40)\,\MeV$,)  but ignoring the
systematic  errors involved in the neglect of the hadronic  contribution,
would be 
\be\lbl{nlla}
\hat{B}_{K}\vert_{\mbox{\rm {\scriptsize nll}}}
\sim 0.96\pm 0.04\qquad\foor\qquad  \muhad= 1.4\,\GeV \,.
\ee 
The problem,  however, with this 
simple estimate is that the underlying assumption of neglecting entirely
the low--energy hadronic integral  does not satisfy any of the long--distance
{\it matching constraints} discussed in the previous section.

One can considerably improve on the previous estimate by taking into
account the contribution from the {\it lowest hadronic state} 
to the low--energy hadronic integral, the $\rho$ vector meson. It turns out
that, in the presence of possible triple poles in Eq.~\rf{largeNansatz}, this
is the {\it minimal hadronic ansatz} approximation at which one can fix the
large--$N_c$ hadronic expression in Eq.~\rf{largeNansatz} to satisfy the two
matching constraints in Eqs.~\rf{Q2} and
\rf{slope} as well as the first non--trivial constraint from the OPE in
Eq.~\rf{OPE4}. In fact, it has been shown that the phenomenological  values
of the
$L_{i}$ constants which appear on the r.h.s. of Eq.~\rf{slope}  agree well 
with those obtained from the integration of the {\it lowest vector}
state~\cite{EGPdeR89,EGLPdeR89,PPdeR98} alone~\footnote{The constant 
$L_3$ gets also an extra contribution from the lowest scalar state, but it
is rather small.}. In this approximation~\cite{PPdeR98},  the
sum rule in  Eq.~\rf{slope} becomes simply
\be
\alpha_{V}\rho_{V}^2+2\beta_{V}\rho_V+3\gamma_V=21\rho_{V}^3\,,\quad\with\quad
\rho_{V}=M_{V}^2/\muhad^2\,.
\ee 
Using this constraint, together with  Eqs. \rf{opeconstraint} 
and \rf{Q2} restricted to
the lowest hadronic state, we can then solve 
for $\alpha_{V}$, $\beta_{V}$ and $\gamma_V$ in terms of
$\rho_{V}$ with the result:
\be
\alpha_{V}=24 \pi \alpha_s\rho_{V} F_0^2/M_{V}^2\,,\quad
\beta_{V}=-\rho_V(3\rho_V +2\alpha_{V})\,,\quad\annd\quad 
\gamma_V=\rho_V^2(9\rho_V+\alpha_{V})\,.
\ee 
The resulting $W[z]$
function, normalized to its value at the origin $W[0]=6$, is plotted 
in Fig.~1 below as a function of $z=Q^2/\muhad^2$. Also shown in the same
plot are  the chiral perturbation behaviour of the $W[z]$ function from
the knowledge of its value and the slope at the origin (the
green line)  and the corresponding behaviour from the OPE expression in
Eq.~\rf{OPE4} (the blue line). We can now make an improved
estimate of
$\hat{B}_{K}$ by calculating the integral 
\be\lbl{sol}
\int_{0}^{\hat{z}} dz\, W[z]+\int_{\hat{z}}^{\Lambda^2/\muhad^2}dz\, 
W_{\mbox{\rm
{\scriptsize OPE}}}[z]\,,
\ee
with $W[z]$ in the first integral approximated by the {\it minimal 
hadronic ansatz} just discussed. The choice
of $\hat{z}$ which minimizes the dependence of $\hat{B}_{K}$ on $\hat{z}$ is
the one at which the two curves $W[z]$ and $W_{\mbox{\rm {\scriptsize
OPE}}}[z]$ intersect. In terms of this $\hat z$, which separates
the long--distance part of the integral estimated with the {\it minimal
hadronic ansatz} and the short--distance part of the integral estimated with
the leading OPE behaviour, the resulting expression for
$\hat B_K$ is
\bea
\lbl{aansatzBK}
&&\hat{B}_{K}  = 
\left(\frac{1}{\als(\muhad)}\right)^{\frac{3}{11}}\frac{3}{4}\left\{1  - 
\frac{\als(\muhad)}{\pi}\left[\frac{1229}{1936}-\frac{3}{4}\log\hat{z}\right]+
\cO\left(\frac{N_c
\als^{2}(\muhad)}{\pi^{2}}\right)\right. \nn\\
&& \left. -\frac{\muhad^2}
{32\pi^2
F_{0}^2}\left[\alpha_{V}\log\frac{\hat{z}+\rho_{V}}{\rho_{V}}
-\beta_{V}\left(\frac{1}{\hat{z}+\rho_{V}}-\frac{1}{\rho_{V}}
\right)- \frac{\gamma_V}{2}
\left(\frac{1}{(\hat z+\rho_V)^2}-\frac{1}{\rho_V^2} \right)\right]\right\}\,.
\eea
As seen in Fig.~1, which corresponds to $\muhad\simeq
1.4\,\GeV$, $F_{0}=85.3\,\MeV$ and $M_{V}=770\,\MeV$, 
the intersection of the hadronic curve and the OPE curve happens at a value
$\hat{z}\simeq 0.39$ (i.e. $Q\simeq 0.88\,\GeV$), at which value 
Eq.~\rf{aansatzBK} gives
\be\lbl{Bkest}
\hat B_{K}\simeq 0.38\,.
\ee
The stability of $\hat B_{K}$ versus $\hat{z}$ is rather good, and it is shown
in Fig.~2 for the same input values as
in Fig.~1.

%%%%%%%%%%%%%%%%%%%%%%%%%%%%%%%%%%% 
%\vspace{0.8cm}
%\centerline{\epsfbox{fignew1U.eps}}
\begin{figure}
\centering
\includegraphics[totalheight=3.5in]{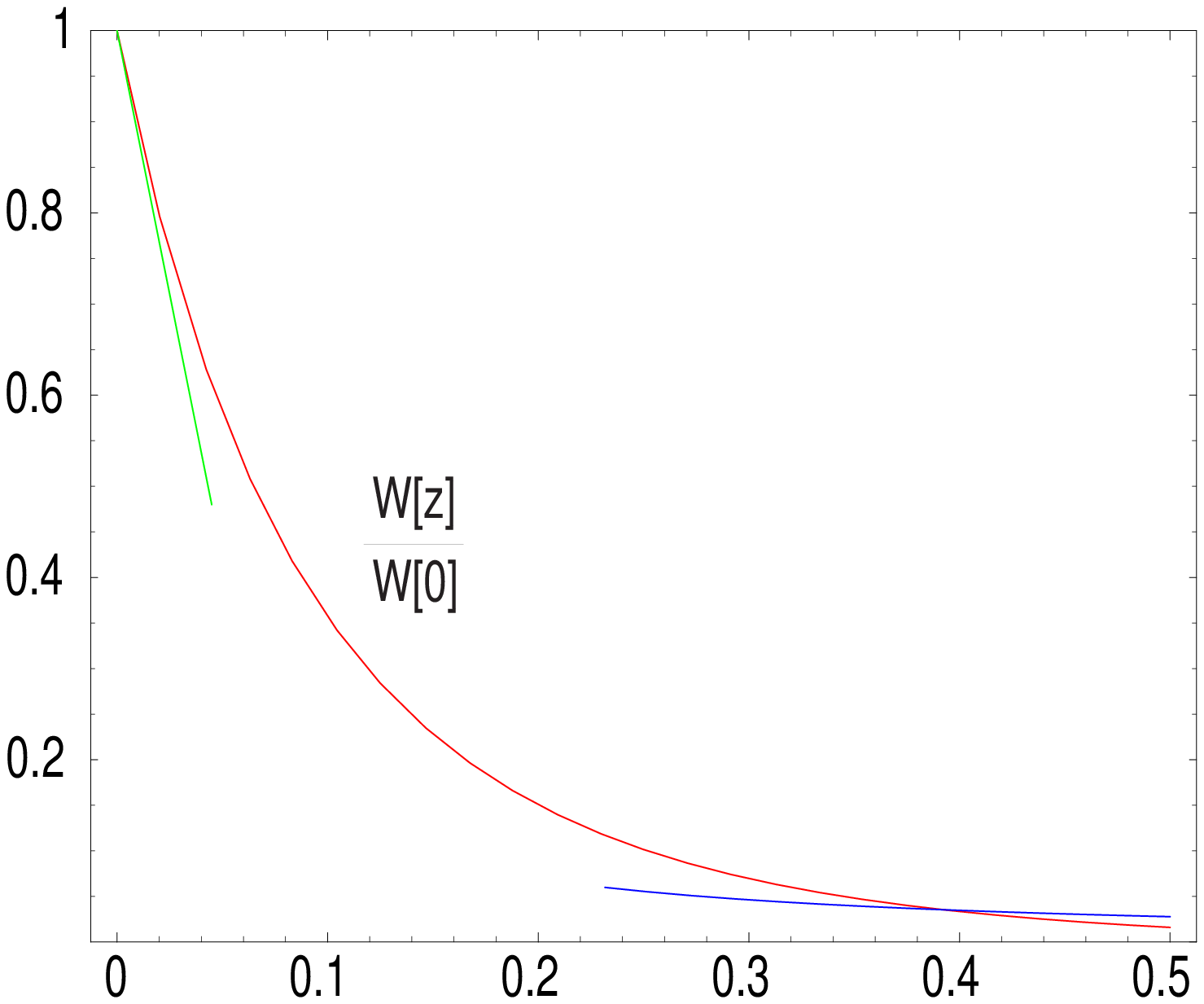}
%\vspace{0.3cm}
%{\bf{Fig.~1}}
\caption{\it Plot of the hadronic function $W[z]$ in 
Eq.~\rf{largeNansatz} versus $z=Q^2/\muhad^2$ for $\muhad=1.4\,\GeV$. 
The red
curve is the normalized function
$W[z]$ corresponding to the {\it minimal hadronic ansatz} discussed in the
text, with a vector meson of mass $M_{V}=770\ \MeV$. The green line
represents the low--energy chiral behaviour and the blue curve the
prediction from the OPE.}
\end{figure}
%\vspace{0.5cm}
%%%%%%%%%%%%%%%%%%%%%%%%%%%%%%%%%%

Notice that, if we had had  a perfect matching between the {\it minimal
hadronic ansatz} and the OPE, we could have taken the limit
$\hat{z}\ra\infty$ in Eq.~\rf{aansatzBK} and find again the same result as in
Eq.~\rf{finalBK} with the sum over hadronic states restricted to the lowest
vector state. The fact that, as seen in Fig.~1, the matching is not perfect
is not surprising; it is due to the restriction of the infinite sum in
Eq.~\rf{largeNansatz} to just the couplings of the lowest vector state. It is
quite remarkable that, already at this approximation, the quality of
the matching is so good. The advocated choice of a
$\hat{z}$ which minimizes the value of
$B_{K}$ at the separation between a low  $Q^2$ region and a high  $Q^2$
region implicitly assumes that the leading term of the OPE controls 
reasonably well  the behaviour of the underlying Green's function $W[z]$
down to $Q$ values in the region $0.8\,\GeV\lesssim Q \lesssim
1\,\GeV$.~\footnote{In principle it is possible to improve on the {\it
minimal hadronic ansatz} approximation we are adopting here by allowing for
couplings of higher states in Eq.~\rf{aansatzBK} 
provided of course that higher order terms in the OPE 
(or the chiral expansion) of the function $W[z]$
are calculated as well. This is under investigation at present.}  

In order to quantify the errors in Eq.~\rf{Bkest} we proceed as follows:
for every choice of
$\muhad$ one finds the corresponding value of $\hat z$ for which the hadronic
ansatz and the OPE intersect. This value of $\hat z$ is then used in
Eq.~\rf{aansatzBK} to obtain
$\hat {B}_{K}$. The error is estimated by allowing for a reasonable variation
of the input parameters: $M_{V}=770 \pm 30\ \MeV$ and $1$~GeV$\leq
\muhad
\leq 3$~GeV. The corresponding values for $\hat{B}_{K}$ are given in
Table~1. They correspond to the spread of values:
\be
\lbl{spread}
0.33\le \hat{B}_{K}\le 0.44 \,.
\ee

%%%%%%%%%%%%%%%%%%%%%%%%%%%%%%%%%%% 
%\vspace{0.8cm}
\begin{figure}
\centering
\includegraphics[totalheight=3.5in]{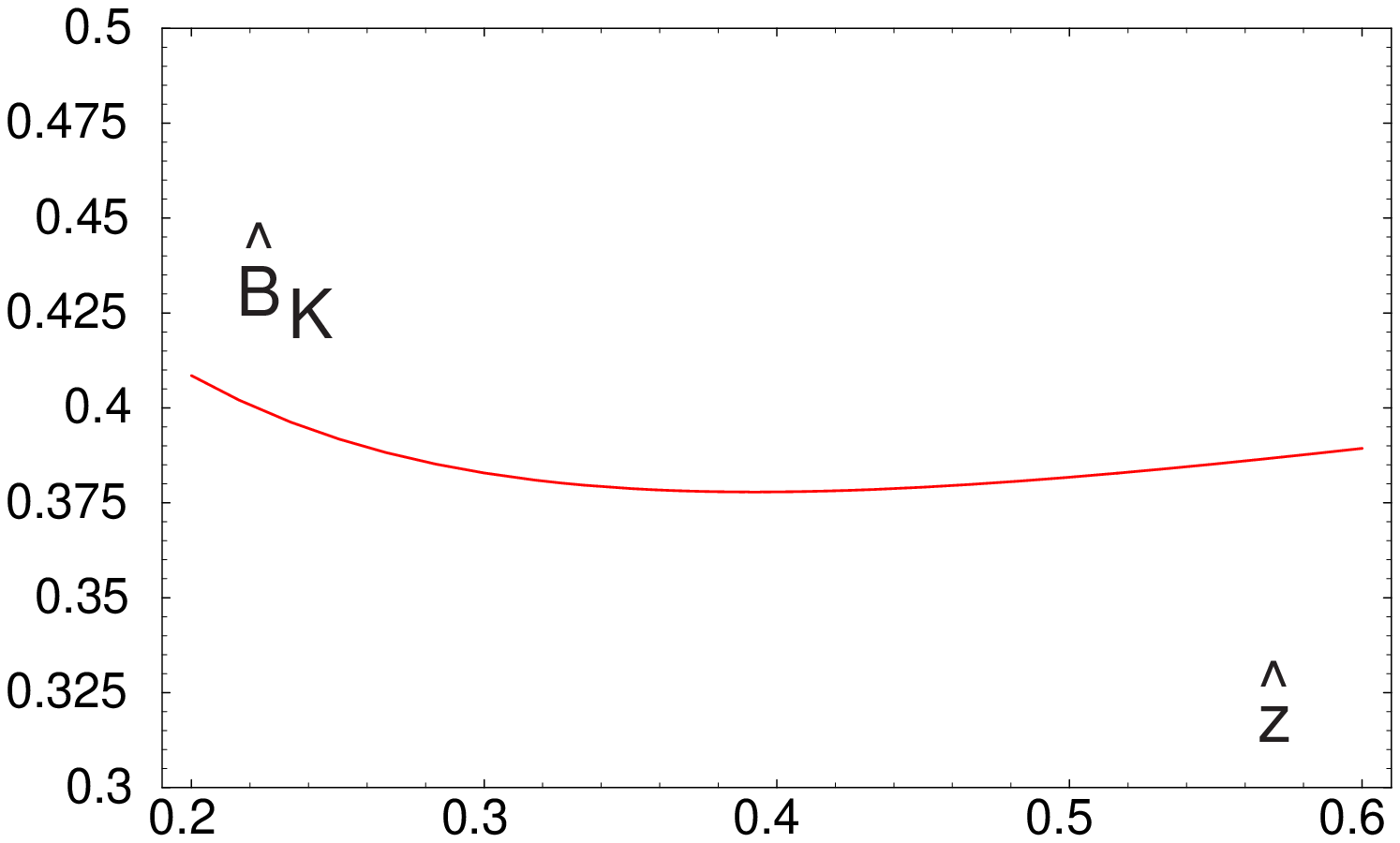}
%\centerline{\epsfbox{fignew2U.eps}}
\vspace{-2cm}
%{\bf{Fig.~2}} 
\caption{\it  Plot of $\hat{B}_{K}$ in Eq.~\rf{aansatzBK} versus
$\hat{z}$, for 
$\muhad\simeq 1.4\,\GeV$, $F_{0}=85.3\,\MeV$ and $M_{V}=770\,\MeV$. Notice
the vertical scale in the figure.}
%\vspace{0.5cm}
\end{figure}
%%%%%%%%%%%%%%%%%%%%%%%%%%%%%%%%%%
%%%%%%%%%%%%%%%%%
\begin{table*}[h]
\caption[Results]{Summary of  $\hat{B}_{K}$ results for different input
values of $\muhad$ and $M_V$}
\lbl{table1}
\begin{center}
\begin{tabular}{||c|c|c|c|c|c|c|c|c|c||}  \hline  {$M_{V}$} &
\multicolumn{9}{c| |}{$\muhad^2$} \\ \cline {2-10} & $1\,\GeV^2$ &
$2\,\GeV^2$  & $3\,\GeV^2$ & $4\,\GeV^2$ & $5\,\GeV^2$ & $6\,\GeV^2$ &
$7\,\GeV^2$ & $8\,\GeV^2$ & $9\,\GeV^2$\\ 
\hline
\hline $740\,\MeV$ & 0.387 & 0.413 & 0.421 & 0.426 & 0.429 & 0.431 & 0.433 &
0.434  & 0.435
\\$770\,\MeV$ & 0.388 & 0.413 & 0.422  & 0.426 & 0.429 &0.432 & 0.433 & 0.435
& 0.436
\\ $800\,\MeV$ & 0.329 & 0.342  & 0.346  &0.347 & 0.348 & 0.349 & 0.350 &
0.350 & 0.350
\\
\hline
\end{tabular}
\end{center}
\end{table*} 
%%%%%%%%%%%%

\noi
One way to quantify the {\it systematic error} of the
{\it minimal hadronic ansatz}  approximation of this calculation would be
to include in the analysis e.g. higher order terms in Eq.~\rf{OPE4}. 
Until we do that we suggest 
as a cautious rule-of-thumb estimate of this uncertainty to round off 
the spread in Eq.~\rf{spread} to an overall $30\%$ of the central value, 
with the result
\be\lbl{finalnumber}
\hat{B}_{K}=0.38\pm 0.11\,.
\ee
Several remarks concerning this result are in order.

\begin{itemize}

\item 
Our result in Eq.~\rf{finalnumber} does not include the error due
to next--to--next--to--leading terms in the $1/N_c$ expansion nor the error
due to chiral corrections in the unfactorized contribution, but we consider
it a rather {\it robust} prediction of $\hat B_K$ in the chiral limit and at
the next-to-leading order in the $1/N_c$ expansion.

\item
Our calculation shows a crucial qualitative issue,
which is the fact that the low--energy hadronic contribution below a mass scale
$1\,\GeV\lesssim\muhad\lesssim 3\,\GeV$ brings down, rather dramatically, the
value of
$\hat{B}_{K}\vert_{\mbox{\rm {\scriptsize nll}}}$ in Eq.~\rf{nlla} evaluated
at that scale. We can already conclude that any calculation which
"ignores" the details of the low--energy hadronic contribution will give an
overestimate of $\hat{B}_{K}$.

\item
The result in Eq.~\rf{finalnumber} is compatible with 
the current algebra prediction~\cite{DGH82} which, to lowest order in
chiral perturbation theory, relates the
$B_{K}$ factor to the $K^{+}\ra \pi^{+}\pi^{0}$ decay rate. In fact, our
calculation of the $B_{K}$ factor can be viewed as a successful prediction of
the $K^{+}\ra \pi^{+}\pi^{0}$ decay rate!  

\item
As discussed in ref.~\cite{PdeR96}
the bosonization of the four--quark operator
$Q_{\Delta S=2}$ and the bosonization of the operator
$Q_{2}-Q_{1}$ which generates $\Delta I=1/2$ transitions  are related to each
other in the combined chiral limit and  next--to--leading order $1/N_c$
expansion. Lowering the value of $\hat{B}_{K}$ from the large--$N_c$
prediction in Eq.~\rf{BKlargeN} to the result in Eq.~\rf{finalnumber} is
correlated with an increase of the coupling constant in the lowest order
effective chiral Lagrangian which generates $\Delta I=1/2$ transitions, and
provides a first step towards a quantitative understanding of the dynamical
origin of the  $\Delta I=1/2$ rule.
 
\item
Finally, we wish to point out that the techniques applied here can be
extended as well to include higher order terms
in the chiral expansion. It remains to be seen if chiral corrections to the
unfactorized term in Eq.~\rf{S=2}  could
be so large,  (of order 100\%?) as to increase our result in
Eq.~\rf{finalnumber} to the values favoured by the lattice QCD
determinations~\footnote{See e.g. the latest review in ref.~\cite{Ku00} and
references therein.} as well as by recent phenomenological
arguments~\cite{Ciuetal00}.

\end{itemize}

\vspace*{7mm} 
{\large{\bf Acknowledgments}}

\vspace*{3mm}

\noi
We wish to thank Marc Knecht, Hans Bijnens and Ximo Prades for very helpful
discussions on many topics related to the work reported here and to Maarten
Golterman,  Michel Perrottet and Toni Pich  for discussions
at various stages of this work.  One of the authors (S.P.) is 
also grateful to C. Bernard and Y. Kuramashi
for conversations  and to the Physics Dept. of Washington University
in  Saint Louis for the hospitality extended to him while the earlier version
of this work was  being finished.
This work has been supported in part by
TMR, EC-Contract No. ERBFMRX-CT980169 (EURODA$\phi$NE). The work of
S.~Peris has also been partially supported by the research project
CICYT-AEN99-0766.

%%%%%%%%%%%%%%%%%%%%%%%%%%
%%%%%%%%%%%%%%%%%%%%%%%%%%
%%%%%%%%%%%%%%%%%%%%%%%%%%
%%%%%%%%%%%%%%%%%%%%%%%%%%
%%%%%%%%%%%%%%%%%%%%%%%%%%

%%%%%%%%%%%%%%%%%%%%%%%%%%%%%%%%%%%%%%%%

\end{document}